\documentclass[aps,prd,reprint,superscriptaddress,preprintnumbers,nofootinbib]{revtex4-1}

\usepackage{amsmath}
\usepackage{amsfonts} 
\usepackage{amssymb}
\usepackage{graphicx}
\usepackage{hyperref}
\usepackage{tensor}
\usepackage{siunitx}
\newcommand{\be}{\begin{equation}}
\newcommand{\ee}{\end{equation}}

\newcommand{\beq}{\begin{equation}}  \newcommand{\eeq}{\end{equation}}
\newcommand{\bal}{\begin{aligned}}   \newcommand{\eal}{\end{aligned}}
\newcommand{\bea}{\begin{eqnarray}}  \newcommand{\eea}{\end{eqnarray}}

\begin{document}

\title{\Large On the cosmological constant, the KK mass scale, and the cut-off dependence in the dark dimension scenario}
\preprint{LMU-ASC 31/23, MPP-2023-191}
\preprint{}

 \author{\large Luis A. Anchordoqui}
\affiliation{Department of Physics and Astronomy, Lehman College, City University of New York, NY 10468, USA}
\affiliation{Department of Physics, Graduate Center, City University of New York, NY 10016, USA}
\affiliation{Department of Astrophysics, American Museum of Natural History, NY 10024, USA}

 \author{\large Ignatios Antoniadis}
\affiliation{Laboratoire de Physique Th\'eorique et Hautes Energies - LPTHE, Sorbonne Universit\'e, CNRS, 4 Place Jussieu, 75005 Paris, France}
\affiliation{Department of Mathematical Sciences, University of Liverpool Liverpool L69 7ZL, United Kingdom}

             \author{\large Dieter L\"ust}
\affiliation{Arnold-Sommerfeld-Center for Theoretical Physics, Ludwig-Maximilians-Universit\"at, 80333 M\"unchen, Germany}
\affiliation{Max-Planck-Institut f\"ur Physik (Werner-Heisenberg-Institut),
            F\"ohringer Ring 6,
             80805, M\"unchen, Germany}
             
             \author{\large Severin L\"ust}
\affiliation{Laboratoire Charles Coulomb (L2C), Universit\'e de Montpellier, CNRS, 
    34095    Montpellier, France}

\begin{abstract}
\vspace{0.5cm}
\noindent In this short note we comment on the relation between the cosmological and the Kaluza-Klein mass scale in the dark dimension scenario~\cite{Montero:2022prj}, also in view of some recent claims~\cite{Branchina:2023ogv}
that would raise some doubts about the validity of this scenario. Here we argue that these claims have serious flaws and cannot be trusted.
\vspace{1cm}
\end{abstract}

\maketitle

The Swampland program aims to distinguish effective field theories (EFT) that can be completed to quantum gravity in the ultraviolet from those which cannot~\cite{Vafa:2005ui}.
In this way it hints towards a new and deep interplay between physics in the UV and in the IR.
One of the swampland conjectures is the anti-de Sitter distance conjecture~\cite{Lust:2019zwm}, which relates the cosmological constant $\Lambda_{cc}$ to the mass scale $m$ of a tower of states:
\begin{equation}\label{adscon}
m \sim \left|\frac{\Lambda_{cc}}{M_p^4}\right|^{\alpha} M_p \;,
\end{equation}
where $\alpha$ is a positive order-one number. This distance conjecture was concretely applied~\cite{Montero:2022prj} to the case of a positive cosmological constant leading to the dark dimension (DD) scenario with one extra dimension of micron size
and a related Kaluza-Klein (KK) mass scale $m\equiv m_{KK}\sim{\cal O}({\rm meV})$, up to a numerical factor. As discussed for example in \cite{Anchordoqui:2022txe,Gonzalo:2022jac,Anchordoqui:2022svl,Anchordoqui:2023oqm},
the DD scenario has many interesting features for particle physics and cosmology. 
However, in a recent work \cite{Branchina:2023ogv} it is claimed that the basic prediction of the DD scenario, namely the relation between  the cosmological constant and the tower mass scale is invalid.
We will discuss that in our opinion the conclusion of~\cite{Branchina:2023ogv} has serious flaws and cannot be trusted.

Let us first recall that the allowed range of the parameter $\alpha$ can be largely restricted. Since the KK tower contains massive spin-two bosons, 
the Higuchi bound~\cite{Higuchi:1986py} provides an absolute upper limit to $\alpha$, namely $\alpha\leq{1/2}$. This bound must be respected in any unitary EFT. 

A lower bound for $\alpha$ follows from explicit calculations of the vacuum energy in string theory (on which we 
will comment further below), and for a $d$-dimensional EFT it is given as $\alpha\geq 1/d$. So in total we have for $d=4$
\begin{equation}
1/4\leq\alpha\leq  1/2\, .
\end{equation}
However, note that this bound on $\alpha$, in particular  $\alpha\leq  1/2$, might be refined in case the relation (\ref{adscon}) contains a large constant on its right hand side.

As explained in \cite{Montero:2022prj}, additional
 experimental arguments, like constraints from 5th-force experiments, then lead to the conclusion that there is one extra dimension of radius $R$ in the micron range, and that the lower bound for $\alpha=1/4$ is basically saturated, i.e. there
is the following parametric  relation between the KK tower mass scale and the cosmological constant (up to another proportionality parameter $\lambda$):
\begin{equation}\label{correct}
\Lambda_{cc} \simeq m_{KK}^4\,.
\end{equation}
Note that for any value of $\alpha$ larger than 1/4, the measured value of $\Lambda_{cc}$ would imply a too small value for $m_{KK}$, incompatible with already existing bounds on the size of the fifth dimension; conversely starting with 
$m_{KK}\sim{\cal O}({\rm meV})$ and increased values of $\alpha>1/4$ leads to too large contributions in the vacuum energy.

Now, it is claimed in~\cite{Branchina:2023ogv} based on a specific EFT loop calculation
that the dominating contribution to the vacuum energy corresponds in the nutshell to $\alpha=1/2$ or to $\alpha=3/2$, 
leading to a too large contribution to $\Lambda_{cc}$. Note that whereas $\alpha=1/2$ just saturates the Higuchi bound, $\alpha=3/2$ parametrically
violates it and therefore is potentially inconsistent with unitarity. However, we will show that with a certain choice of the UV cuf-off the Higuchi bound can be rescued. 
In other words, it will depend on the choice of the UV cut-off whether the Higuchi bound is violated or not.

Second, as already indicated, their results contain some particular UV cut-off dependence, 
 where it all depends on how one regularizes the infinite integral.
This makes the EFT result a priori
ill defined and therefore the EFT calculation is potentially highly misleading. In particular, it also does not take into account quantum gravity effects, like UV-IR mixing, which we will mention later.
In other words, 
without fixing the value of the cut-off in their formulae, it is meaningless to extract any relation between  $\Lambda_{cc}$ and $m_{KK}$. 
On the other hand, 
in string theory or in quantum gravity, the cut-off is removed and the string computations are well defined and have been confirmed 
(see e.g.~\cite{Itoyama:1986ei,Itoyama:1987rc,Antoniadis:1991kh,Bonnefoy:2018tcp}) several times leading to a finite result behaving as in eq.(\ref{correct}).

In field theory, a particular example is QFT at finite temperature leading to the same behaviour $T^4$ where $T=1/R$. No string regularisation is used for that. Actually, the finite temperature example is very similar to the Casimir energy computation which can be computed with completely convergent diagrams yielding the result~\eqref{correct} (see for instance~\cite{Arkani-Hamed:2007ryu} for any $d$). Note that in both cases, the result should vanish at the decompactification limit or at $T=0$, which is automatic if the higher-dimensional theory becomes supersymmetric; it is also a condition imposed by the swampland distance conjecture~\cite{Ooguri:2006in}.
It is worth pointing out that in the string calculations modular invariance of the one-loop amplitudes plays a decisive role, which dictates a particular regularisation for the associated EFT computations. It amounts to perform first a Poisson resummation and then subtract the infinite zero mode contribution which is radius/temperature independent. 

Let us finally have a more close look into the result of~\cite{Branchina:2023ogv}. Their EFT calculation leads to the following two relations (see eqs.(26) and (27) in \cite{Branchina:2023ogv}), and, depending how supersymmetry is broken,
the EFT relations are (up a log correction in the second case):
\begin{equation}\label{relations}
\Lambda_{cc}\sim m_{KK}^2R\Lambda^3 \quad {\rm or} \quad\Lambda_{cc}\sim m_{KK}^{2/3}R^{5/3}\Lambda^5\,.
\end{equation}
The first case would correspond to $\alpha=1/2$ and the second case to the "forbidden" value $\alpha=3/2$.
However, 
these formulas still contain the radius $R$ of the 5th dimension and the UV cut-off $\Lambda$. The Higuchi bound will now put a bound on $R$ and $\Lambda$.
Specifically, the Higuchi bound is satisfied in case the following relations hold for the two cases:
\begin{equation}\label{Hbound}
R\Lambda^3 \leq M_P^2\quad {\rm or} \quad R\Lambda^3 \leq (\Lambda_{cc}M_P)^{2/5}\simeq10^{-48}M_P^2\, .
\end{equation}
In the first case $R\Lambda^3$ is not strongly restricted, but for the second case due to the smallness of $\Lambda_{cc}\simeq 10^{-120}M_P^4 $ a very low cut-off is required by the Higuchi bound.

Finally, we will show how the relations (\ref{relations}) can be made consistent 
with the DD scenario by a certain, natural choice for $R$ and $\Lambda$. First these formulas still contain the radius $R$ of the 5th dimension,
which should be replaced by $R=1/m_{KK}$. Hence additional powers of $m_{KK}$ will show up in the result.

Secondly, the UV cut-off $\Lambda$ is not a constant but crucially depends also on the IR mass scale, given in terms of $\Lambda_{cc}$. This is precisely a manifestation of the before mentioned 
UV-IR mixing in quantum gravity or in string theory, which will eliminate the arbitrariness from the EFT calculation. Therefore, one has to determine what is the correct cut-off. Actually, the largest possible cut-off
in any EFT coupled to quantum gravity is the species scale $\Lambda_{sp}$ \cite{Dvali:2007hz}. This is the scale where gravity becomes strongly coupled and the EFT
necessarily breaks down. For the DD scenario, it has the following dependence on $m_{KK}$: $\Lambda=\Lambda_{sp}\simeq m_{KK}^{1/3}M_p^{2/3}$. However, inserting $\Lambda_{sp}$ into eq.(\ref{relations}) does not  reproduce the string calculation. Specifically, one gets that 
\begin{equation}
\Lambda_{cc}\sim m_{KK}^2M_P^2 \quad {\rm or} \quad\Lambda_{cc}\sim m_{KK}^{2/3}M_P^{10/3}\,.
\end{equation}
The second relation now explicitly violates the Higuchi bound.

An alternative  choice for  the cut-off is to identify it with the KK
scale, i.e. $\Lambda=\Lambda_{KK}\simeq m_{KK}$. This is the scale
where the 4-dimensional EFT description breaks down and turns into a
5-dimensional EFT, at least for what the gravity interactions
in the 5D bulk concern. This choice of the cut-off is also in-line with the corresponding
string calculation, where only the lowest KK mode effectively contributes.  Plugging this cut-off into both relations of eq.(\ref{relations}), one immediately sees that the swampland relation (\ref{correct})
is indeed satisfied for both cases.

In summary, in this note we argue that there is 
no reason to conclude  that the DD scenario is invalid. On the contrary, we 
pointed out the caveats of the  EFT calculation in~\cite{Branchina:2023ogv} with respect to the 
cut-off dependence and showed how, by the  choice of a particular cut-off, it is consistent with the corresponding string calculation and swampland prediction. Indeed, the swampland guiding principles point to $\Lambda_{cc} \sim m_{KK}^{1/\alpha}$ while theoretical and experimental bounds 
single out the power $m_{KK}^4$, implying that the fundamental nature of the DD is not just a matter of summing EFT contributions.\footnote{This practical reasoning has also been missed in~\cite{Burgess:2023pnk}.}

\vspace{10px}
{\bf Acknowledgements}
\noindent

We thank Ivano Basile, Miguel Montero, Cumrun Vafa and Irene Valenzuela for useful 
discussions. The work of D.L. is supported  the Origins Excellence Cluster. S.L. thanks the Max-Planck-Institut f\"ur Physik for hospitality during the completion of this work.
L.A.A. is supported by the US National Science Foundation PHY-2112527.


\end{document}